\documentclass[traditabstract,longauth]{aa}
\usepackage{txfonts}
\usepackage{graphicx}

\usepackage{stfloats}
\usepackage{natbib}
\usepackage{subfigure}
\usepackage{arydshln}
\bibpunct{(}{)}{;}{a}{}{,} 

\newcommand\1{{\scriptsize I}}
\newcommand\2{{\scriptsize II}}
\newcommand\3{{\scriptsize III}}
\newcommand\4{{\scriptsize IV}}

\begin{document} 

\title{The VLT-FLAMES Tarantula Survey\\ II: R139 revealed as a massive binary system }

\titlerunning{R139 revealed as a massive binary system}
\author{W.~D.~Taylor \inst{1}
\and C.~J.~Evans \inst{2,1}
\and H.~Sana \inst{3}
\and N.~R.~Walborn \inst{4}
\and S.~E.~de Mink \inst{4}\thanks{Hubble Fellow}
\and V.~E.~Stroud \inst{5,6}
\and A.~Alvarez-Candal\inst{7}
\and R.~H.~Barb\'{a}\inst{8,9}
\and J.~M.~Bestenlehner \inst{10}
\and A.~Z.~Bonanos \inst{11}
\and I.~Brott \inst{12,13}
\and P.~A.~Crowther  \inst{14}
\and A.~de~Koter \inst{3,12}
\and K. Friedrich \inst{15}
\and G.~Gr$\ddot{\textup{a}}$fener  \inst{10}
\and V.~H\'{e}nault-Brunet \inst{1}
\and A.~Herrero \inst{16,17}
\and L.~Kaper\inst{3}
\and N.~Langer \inst{15}
\and D.~J.~Lennon \inst{4}\thanks{European Space Agency}
\and J.~Ma\'{i}z~Apell\'{a}niz \inst{18}
\and N.~Markova\inst{19}
\and N.~Morrell\inst{20}
\and L.~Monaco\inst{7}
\and J.~S.~Vink  \inst{10}
} 

\institute{Institute for Astronomy, Royal Observatory Edinburgh, Blackford Hill, Edinburgh. EH9 3HJ, UK
\and %2
UK Astronomy Technology Centre, Royal Observatory Edinburgh, Blackford Hill, Edinburgh. EH9 3HJ, UK
\and %3
Astronomical Institute Anton Pannekoek, University of Amsterdam, Science Park 904, 1098XH Amsterdam, The Netherlands
\and %4
Space Telescope Science Institute, 3700 San Martin Drive, Baltimore, MD 21218, USA
\and %5
Department of Physics and Astronomy, The Open University, Walton Hall, Milton Keynes, MK7 6AA, UK
\and %6
Faulkes Telescope Project, University of Glamorgan, Pontypridd, CF37 1DL, Wales, UK
\and %7
European Southern Observatory, Alonso de Cordova 3107, Casilla 19001, Santiago 19, Chile
\and %7
Departamento de F$\acute{\i}$sica, Universidad de La Serena, Cisternas 1200 Norte, La Serena, Chile
\and
Instituto de Ciencias Astronomicas, de la Tierra, y del Espacio(ICATE-CONICET), Av. Espana 1512 Sur, 5400 San Juan, Argentina 
\and %8      
Armagh Observatory, College Hill, Armagh, BT61 9DG, Northern Ireland, UK
\and %9
Institute of Astronomy \& Astrophysics, National Observatory of Athens, I. Metaxa \& Vas. Pavlou Street, P. Penteli 15236, Greece
\and %10
Astronomical Institute, Utrecht University, Princetonplein 5, 3584 CC, Utrecht, The Netherlands
\and %11
University of Vienna, Department of Astronomy, T\"{u}rkenschanzstr. 17, A-1180 Vienna, Austria
\and %12
Dept. of Physics \& Astronomy, Hounsfield Road, University of Sheffield, S3 7RH, UK
\and %13
Argelander-Unstitut fur Astronomie der Universitat Bonn, Auf dem Hugel 71, 53121 Bonn, Germany
%\and %14
%Scottish Universities Physics Alliance (SUPA), Institute for Astronomy, University of Edinburgh, Edinburgh, EH9 3HJ, UK
\and %13
Instituto de Astrof$\acute{\i}$sica de Canarias, E-38200 La Laguna, Tenerife, Spain
\and %14
Departamento de Astrof$\acute{\i}$sica, Universidad de La Laguna, Astrof$\acute{\i}$sico Francisco S\'{a}nchez, E-38071 La Laguna, Tenerife, Spain
\and %15
Instituto de Astrof\'{i}sica de Andaluc\'{i}a-CSIC, Glorieta de la Astronom\'{\i}a s/n, E-18008 Granada, Spain
\and
Institute of Astronomy with National Astronomical Observatory, Bulgarian Academy of Sciences, PO Box 136, Smoljan, Bulgaria
\and
Las Campanas Observatory, Carnegie Observatories, Casilla 601, La Serena, Chile.
}

\date{Received February 2011 / Accepted March 2011}
 
\abstract{We report the discovery that R139 in 30 Doradus is a massive spectroscopic binary system. Multi-epoch optical spectroscopy of R139 was obtained as part of the VLT-FLAMES Tarantula Survey, revealing a double-lined system. The two components are of similar spectral types; the primary exhibits strong C~\3 $\lambda 4650$ emission and is classified as an O6.5\,Iafc supergiant, while the secondary is an O6\,Iaf supergiant. The radial-velocity variations indicate a highly eccentric orbit with a period of 153.9 days. Photometry obtained with the Faulkes Telescope South shows no evidence for significant variability within an 18 month period. The orbital solution yields lower mass limits for the components of $M_1 \sin^3{i} = 78 \pm 8\,M_{\sun}$ and $M_2 \sin^3{i} = 66 \pm 7\,M_{\sun}$. As R139 appears to be the most massive binary system known to contain two evolved Of supergiants, it will provide an excellent test for atmospheric and evolutionary models.
}
   
\keywords{binaries: spectroscopic -- stars: early-type -- stars: individual: R139 -- open clusters: individual: 30 Doradus}

\maketitle

%==========================================================================
%
\section{Introduction}
Massive binary stars provide vital insights to our understanding of massive-star evolution. This is primarily due to the accuracy with which their masses can be determined: an essential ingredient for understanding a wide range of stellar properties. Because of the additional constraints that can be placed on their age and evolution, these stars provide information on initial masses, chemical mixing and mass-loss \citep{moffat_binaries,selma}. In a broader context, they can then act as crucial calibration points for models of both stellar atmospheres and evolution.
 
From the catalogue of bright stars in the Magellanic Clouds by \citet{feast}, R139 has a $V$-band magnitude of $\sim$12, making it one of the brightest objects in the 30 Doradus nebula\footnote{Other aliases of R139 include: Brey~86 \citep{brey}, Parker~952 \citep{parker}, BAT99-107 \citep{batt}, and Selman~2 \citep{selman}.}. \citet{walborn_and_b} noted R139 as potentially one of the most massive stars in 30 Doradus, urging more detailed study. 

Multi-epoch spectroscopy of R139 was obtained as part of the campaign by \citet{moffat}. His mean radial velocity from observations in 1982 showed an offset of $\sim$100\,kms$^{-1}$ compared to the mean velocity from earlier data. He noted R139 as a single-lined binary, with a tentative period of
52.7\,d adopted from a number of possible fits to the data.  \citet{schnurr} presented spectroscopy of R139 from three observing seasons (spanning 2001 to 2003). While noting that the system displayed `slightly variable radial velocity', it was concluded that R139 was single, citing the relatively large uncertainties in Moffat's past work for the conflicting scenarios.

R139 has now been observed as part of the VLT-FLAMES Tarantula Survey\footnote{Observations obtained at the European Southern Observatory Very Large Telescope in programme 182.D-0222.} (VFTS), an ESO Large Programme which has obtained multi-epoch spectroscopy of over 800 massive stars in 30~Doradus. A full overview of the survey is given by Evans et al. (Paper~I, 2011), here we report on the discovery of R139 as a massive double-lined binary.

%==========================================================================
%

\section{Observations}\label{observations}
The observations of R139 are summarised in Table \ref{tab_1}. The primary dataset of the VTFS has been obtained with the Giraffe spectrograph using the `Medusa' fibre-fed mode of the FLAMES instrument on the Very Large Telescope (VLT). Details of the reductions and observational strategy can be found in Paper~I, in which R139 is catalogued as object VTFS\,527.

After the initial detection of binarity made from the FLAMES data, follow-up observations have been obtained on the 6.5m Magellan Clay Telescope with the MagE instrument, with X-Shooter on the VLT and also with FEROS on the MPG/ESO 2.2m telescope at La Silla\footnote{ FEROS observations obtained as part of programme 086.D-0997}. All these follow-up observations provide coverage across the entire visible spectrum, with a typical signal-to-noise ratio of order 150 - although the FEROS data must be degraded to $R\sim 9000$ to achieve this.

%==========================================================================
% TABLE OF OBSERVATIONS
\begin{table}[t]
\caption{Observational epochs for R139 detailing; the instrument used, the resolving power ($R$), the date and the time elapsed since the first epoch ($\Delta $ HJD). 
\label{tab_1} } 
\tiny
\begin{center} 
\begin{tabular}{llccr}
\hline\hline 
Epoch & Instrument\,/\,Setting & $R$  & HJD +2400000 & $\Delta $ HJD \\

\hline
1 & FLAMES\,/\,LR02 &  7,000  & \phantom{it}54748.7769\tablefootmark{a} & - \\
2 & FLAMES\,/\,LR02 &   & 54748.8228 & 0.046\\
3 & FLAMES\,/\,LR02 & & 54749.7233 & 0.946\\
4 & FLAMES\,/\,HR15N & 16,000  & 54749.7726 & 0.996\\
5 & FLAMES\,/\,LR03 &  8,500  & 54755.6882 & 6.911\\
6 & FLAMES\,/\,HR15N &    & 54810.6697 & 61.893\\
7 & FLAMES\,/\,LR03 &   & 54810.7374 & 61.961\\
8 & FLAMES\,/\,LR03 &  & 54810.7808 & 62.004\\
9 & FLAMES\,/\,LR03 &  & 54810.8357 & 62.059 \\
10 & FLAMES\,/\,LR02 &   & 54837.6332 & 88.856\\
11 & FLAMES\,/\,LR02 &  & 54868.5569 & 119.780\\
12 & FLAMES\,/\,LR02 &  & 55112.8129 & 364.036\\
\hdashline
13 & MagE \,/\,1'' slit &  4,500 & 55527.8017 & 779.025\\
\hdashline
14 & X-Shooter\,/\,0.5'' slit & \phantom{it}9,100\tablefootmark{b} & 55580.5477  & 831.771\\
15 & X-Shooter &  & 55584.7076 & 835.931\\
16 & X-Shooter  &  & 55588.6922  & 839.915\\
\hdashline
17 & FEROS & 48,000 & 55604.6909  & 855.936 \\
18 & FEROS  &  & 55605.5613 & 856.739\\

\hline
\vspace{-0.6cm}
\end{tabular} 
\end{center} 
\tablefoottext{a}{The FLAMES observations were comprised of back-to-back observations so only the mid-exposure HJD is given.}\ \newline
\tablefoottext{b}{The resolving power quoted for X-shooter is that of the UV-arm only, which overlaps the LR02-03 regions of the FLAMES data.}
\vspace{-0.3cm}
\end{table}

%==========================================================================
%
\section{Results}\label{results}
%==========================================================================
%DETECTION
\subsection{Binary identification and spectral classification}
\citet{feast} described the spectrum of R139 as exhibiting `O-type absorption plus weak W emission'. \citet{walborn_and_b} argued that R139 has very strong Of emission features, classifying it as O7 Iafp, while Schnurr et al. (2008b) adopted the spectral type WN9h::a. This ambiguity has likely arisen because the emission features are the superposition from two similar stars. When observed at lower resolution, this would have falsely suggested enhanced emission or broadening of the lines.

\begin{figure}[!b]
\centering
\includegraphics[width=9.0cm]{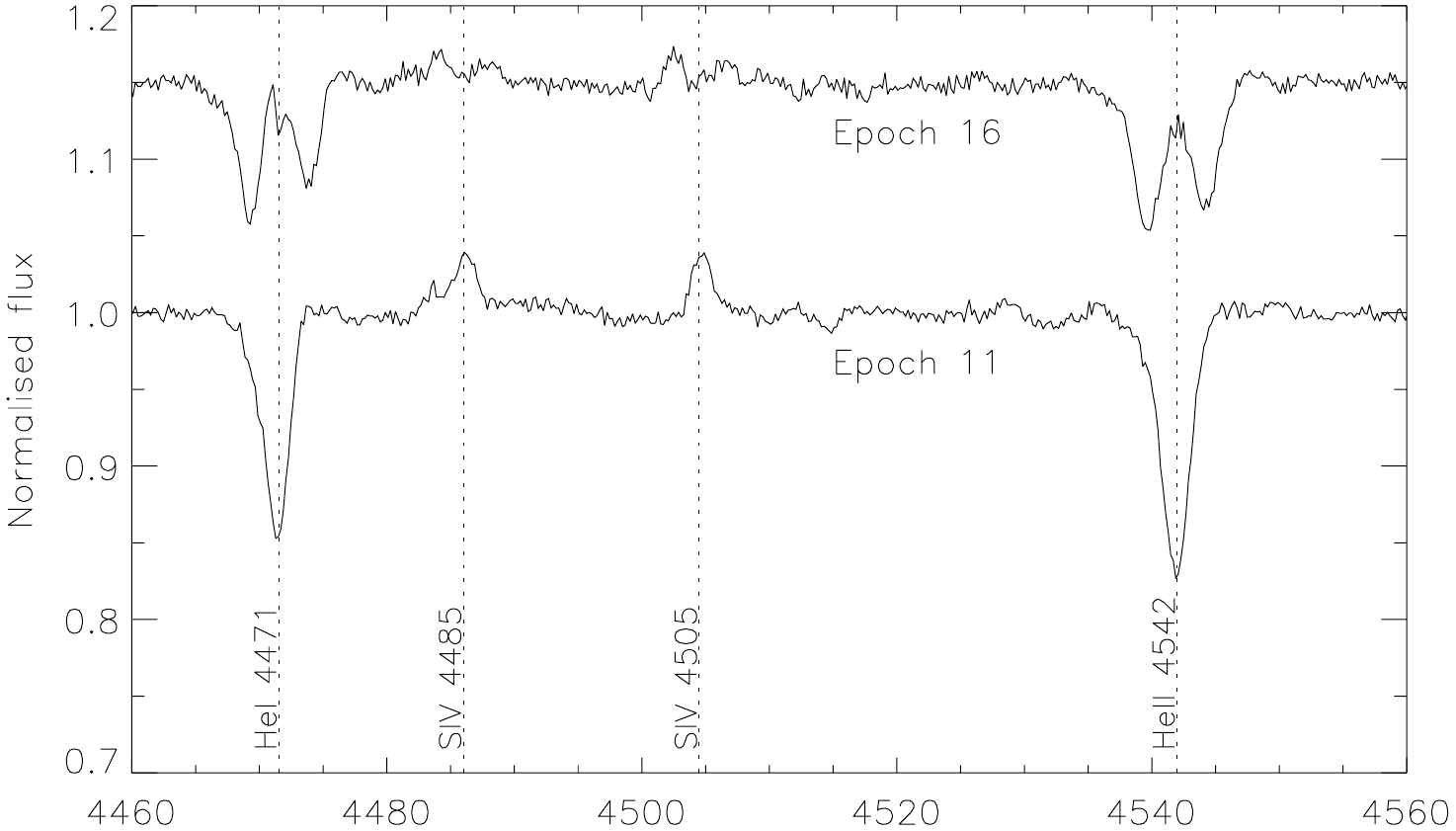}
\caption{Normalised spectra of R139 at epochs of minimum (\#11) and maximum (\#16) observed separation. The He~\1 $\lambda$4471 line profile in epoch \#16 suffers from a slight nebular over-subtraction but does not prevent clear identification of the components.
\label{spectra}}
\vspace{-0.4cm}
\end{figure}

The increased resolution and time-sampling of the new data have revealed that many of R139's prominent emission and absorption features separate into two distinct and similar components. This is best illustrated by contrasting epochs 11 and 16 in Figure~\ref{spectra}, which show the minimum and maximum observed separations respectively. The epochs which display well-separated components have allowed a precise classification of both components. The system consists of a more massive and more luminous primary, which is an O6.5\,Iafc supergiant, and a slightly less luminous O6\,Iaf companion. These spectral types are determined from visual inspection of the He~\2 / He~\1 absorption line ratios between $\lambda$4200 / $\lambda$4026 and also between $\lambda$4542 / $\lambda$4471, the latter of which can be seen in Figure~\ref{spectra}. 

\begin{figure}[t]
\centering
\includegraphics[width=8.cm,height=4.6cm]{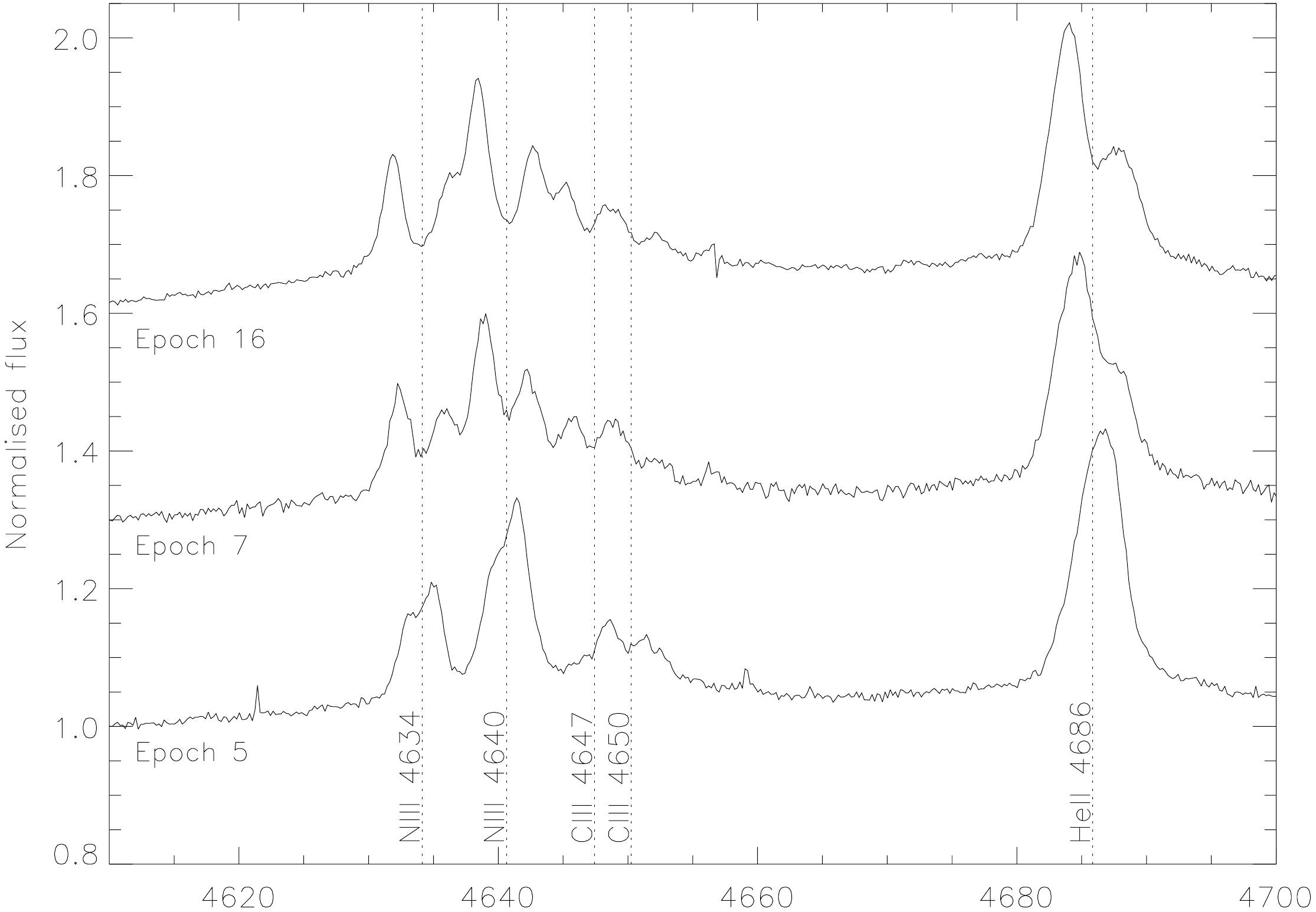}
\caption{Normalised spectra showing the complex N~\3 and C~\3 emission region. Both of the N~\3 emission lines and the He~\2 $\lambda$4686 emission are comprised of two un-equal components, the stronger of which is associated with the primary. The C~\3 emission lines show no separation, but exhibit radial velocity shifts in the same direction as the primary.
\label{carbon}}

\end{figure}

The Ifc classification of the primary arises from emission of C~\3 $\lambda$4647 and  $\lambda$4650-4652. This relatively rare feature led to the recent introduction of the Ofc category in the morphological framework used to classify O-type spectra \citep{walborn_ofc}. Figure~\ref{carbon} clearly illustrates that the N~\3 emission lines separate into two distinct components, whereas the two C~\3 emission lines show no evidence of separation, but merely shift between epochs in the same sense as the primary. The Ofc feature has previously been associated with the O5 spectral type \citep{walborn_ofc}; this therefore, is an interesting example of the phenomenon in a later-type star, albeit in the LMC. 

Where possible, the relative shift of the components was identified through the C~\3 emission and also He~\1 $\lambda$4922 emission, which are only present in the primary. In the LR02 observations, the Si~\4 $\lambda$4116 emission line was used as the main diagnostic for the relative shift of the components. The Struve-Sahade effect \citep[e.g.][]{linder} can most likely be neglected in this system given its relatively long period.

%==========================================================================
%CARBON FIGURE

%==========================================================================
%ORBITS
\subsection{Radial velocity analysis and lower mass limits}
A global $\chi^2$ fitting approach has been used to determine the radial velocity shifts of the different epochs. This technique fits double Gaussian profiles to a number of lines: He~\1 $\lambda$4026, Si~\4 $\lambda$4116, He~\2 $\lambda$4200, S~\4 $\lambda$4485, S~\4 $\lambda$4505, He~\2 $\lambda$4542, He~\2 $\lambda$ 4686 and He~\1 $\lambda$4922. The fitting is performed simultaneously on all the observations, which ensures that consistent profile shapes are used, including at conjunction. This approach improves the disentangling of the contribution from each star for the data sets with limited phase coverage, but ignores the possibility of line profile variations. The formal errors on the measurements for each component are a few km\,s$^{-1}$.

The mass ratio of the system can be found from the ratios of the primary and secondary radial velocities, and is independent of any other assumptions about the orbit \citep{rauw2000}. For R139 the mass ratio is found to be: $M_1 / M_2 = 1.20 \, \pm \, 0.05$.

\begin{figure}[t]
\begin{center}
\includegraphics[width=7.5cm]{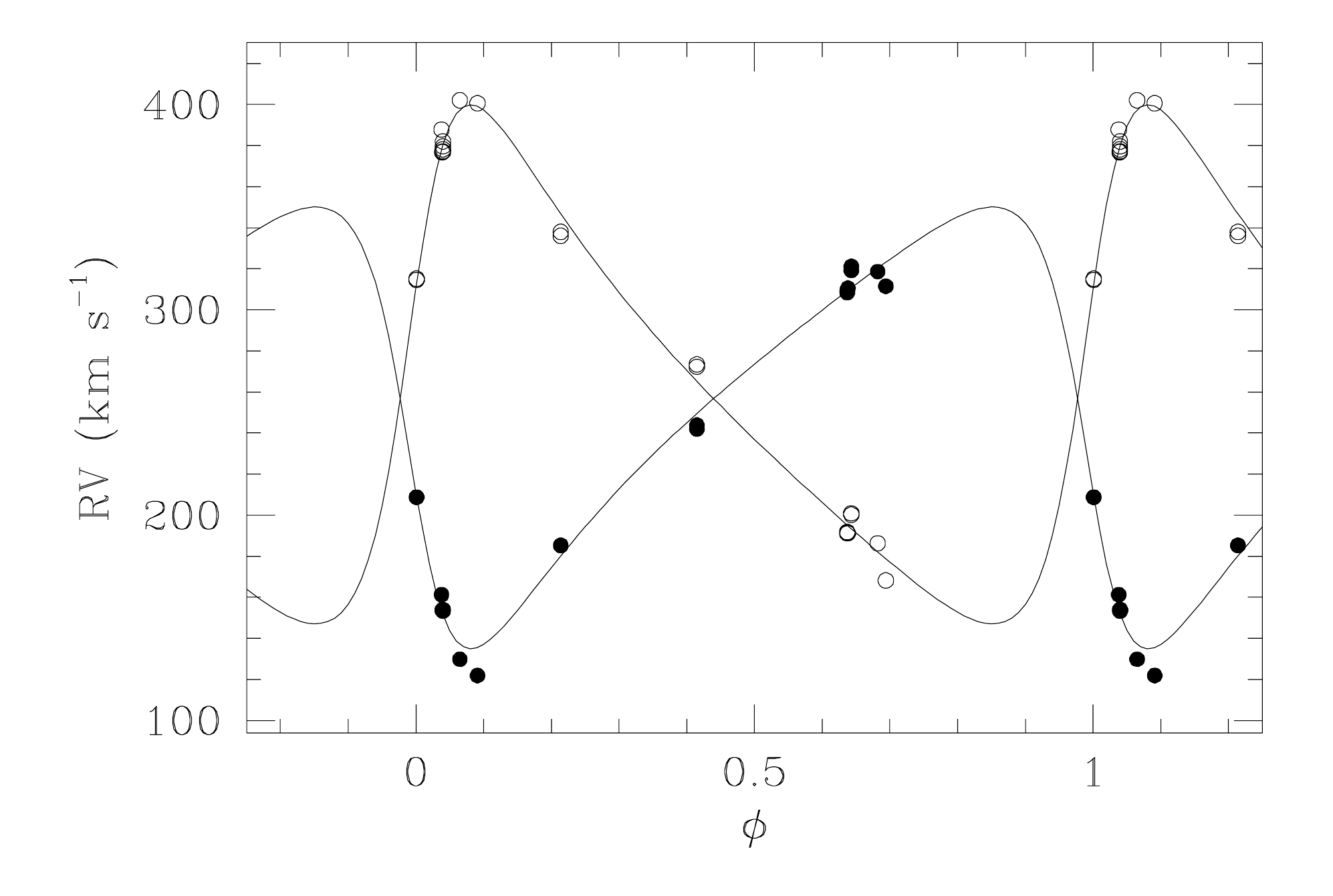}
\caption{The best-fit orbital solution from the measured radial velocities of the components, indicating a 153.9 day orbit. The closed circles denote the velocities of the primary and the open circles the secondary.}\label{orbit}
\end{center}
\vspace{-0.4cm}
\end{figure}

Period searches based on Fourier analysis of the measured radial velocity shifts were performed using the methods of \cite{gosset}: the dominant signal indicated a period of 153.9 days. The corresponding orbital solution, which has an rms uncertainty of 6.2\,km\,s$^{-1}$, is shown in Figure \ref{orbit} \citep[based on methods of][]{sana06stars}. The corresponding lower mass limits for the stars are: $M_1 \sin^3{i} = 78 \pm 8\,M_{\sun}$ and $M_2 \sin^3{i} = 66 \pm 7\,M_{\sun}$. Other orbital parameters are listed in Table \ref{orb_params}. These are sensitive to the behaviour of the system around periastron: further observations of this stage in the orbit would allow confirmation of these results.

\begin{table}[b]
\caption{The parameters associated with the best-fit orbital solution. The errors quoted are the formal errors on the best-fit from the Fourier analysis, and therefore may not be fully representative of the uncertainty in the parameter values.
\label{orb_params} } 
\tiny
\begin{center} 
\begin{tabular*}{0.45\textwidth}{@{\extracolsep{\fill}}ll}
\hline\hline 

Property & Best-fit value \\

\hline

Period, P  & 153.9  $\pm$   0.1\,days\\
Eccentricity, e &  0.46       $\pm$   0.02\\
Argument of periastron, $\omega$  &  106.9       $\pm$   5.0\,deg\\
Date of $\phi=0$ (HJD - 2450000), T$_0 $ &    6035.9    $\pm$   1.3\,days\\
Maximum velocity of primary, K1 & 107.8       $\pm$   3.8\,km\,s$^{-1}$ \\
Maximum velocity of secondary, K2 &   127.0       $\pm$   4.5\,km\,s$^{-1}$\\
Projected semi-major axis for primary, a$_1 \sin{i}$ &  290.6       $\pm$  10.8\,$R_{\sun}$\\
Projected semi-major axis for secondary, a$_2 \sin{i}$ &  342.3       $\pm$  12.8\,$R_{\sun}$\\

\hline
\vspace{-0.6cm}
\end{tabular*} 
\end{center} 
\end{table}

%==========================================================================
%LUMINOSITY
\subsection{The luminosity of R139}
To estimate the luminosity of the system, model atmospheres were calculated with CMFGEN \citep{hillier}, adopting abundances from \cite{asplund} and scaling them appropriately for the LMC. These were used to constrain the effective temperature ($T_{\rm eff}$) consistent with: 
1) the absence of N~\4 $\lambda$4058 emission; 
2) the presence of N~\3 $\lambda$4640 emission;
3) the intensity of the He~\1 $\lambda$4471 absorption.  
This gives an estimated $T_{\rm eff}$ for both components of $34 \pm 2$\,kK.

The luminosity was then determined by matching optical and infrared photometry from \cite{selman} and 2MASS \citep{skrutskie}. For this purpose, the visual extinction $A_V=R\times E(B-V)$ was determined for each model, based on the relation $R = 1.12 \times E(V-K)/E(B-V) + 0.02$ from \citet{fitzpatrick}. The resulting luminosity for the composite system is $\log(L/L_{\odot})=6.4\pm 0.1$ (with $R$ in the range 3.4--3.9).  

%==========================================================================
%

\section{Photometric Variability}\label{fts}
R139 was identified as showing slight photometric variability by \cite{feit}. They observed a 0.3 magnitude dimming in the $V$ band over a 25 day period. However, this was from only three observations taken with a wide (18'') aperture.
  
An active component of the VTFS is photometric follow-up with the 2\,m Faulkes Telescope South, which has been used to monitor seven fields in the 30~Dor region. The default mode of the camera is 2\,$\times$\,2 binning of the CCD pixels, giving an effective pixel-scale on the sky of 0\farcs278. 

\begin{figure}[t]
\begin{center}
\includegraphics[width=8.0cm,height=5.1cm]{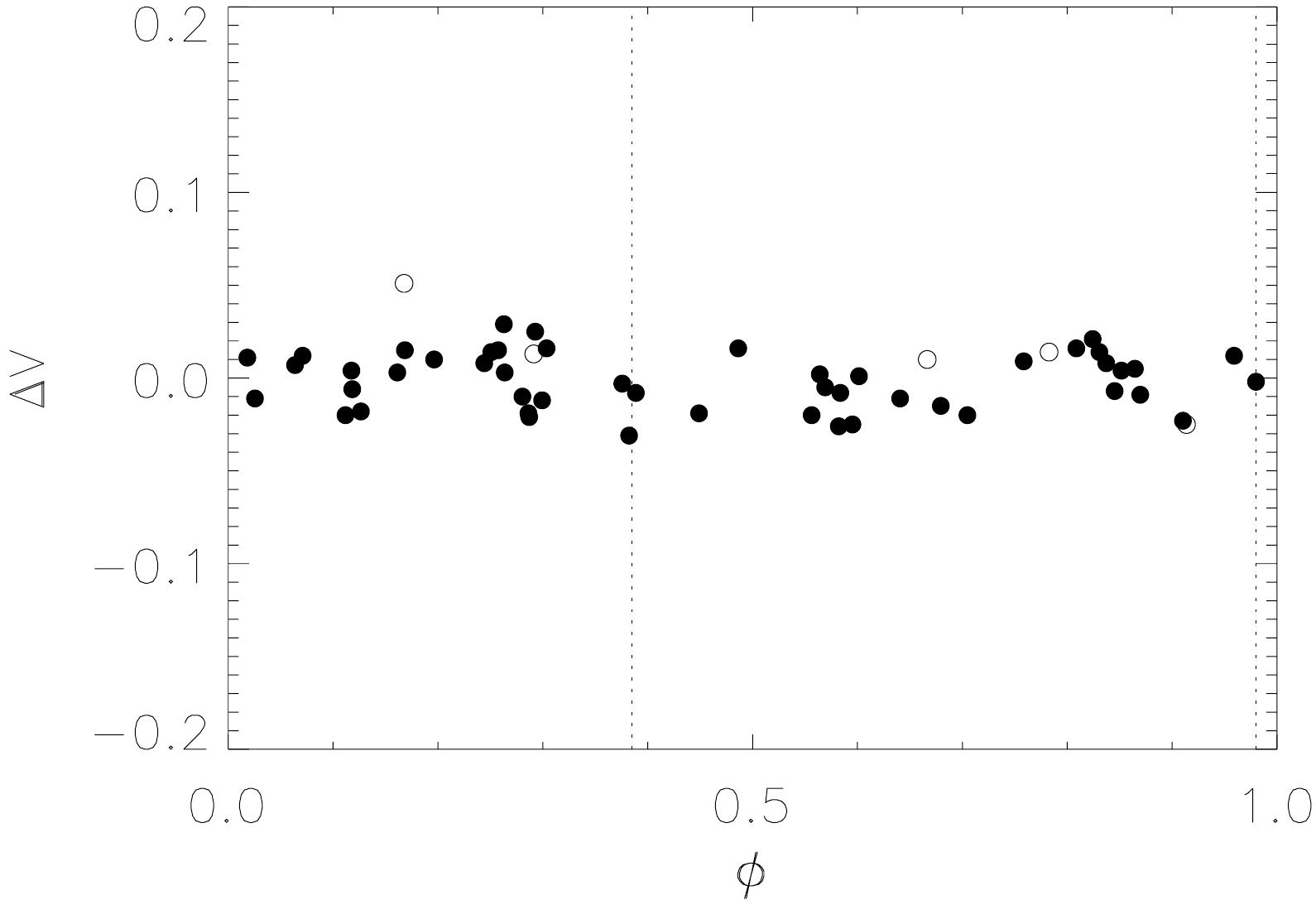}
\caption{Differential V-band residuals for R139 compared to the mean of the five check stars, phased to the 153.8\,day orbit. Open circles denote the five epochs when the seeing was in excess of 2\farcs0. The dotted lines indicate the expected mid-point of any possible eclipses: $\phi = 0.39$ and 0.98 for the eclipses near apastron and periastron respectively.  
}\label{phot}
\end{center}
\vspace{-0.4cm}
\end{figure}

We have 54 V-band epochs for the relevant field, spanning an 18 month period starting in January 2009. The Faulkes data are reduced automatically following observations, but are not calibrated photometrically. Given the crowding in this field, we used {\sc apphot} in {\sc iraf} to obtain instrumental magnitudes of R139 from aperture photometry. Five `check' stars of similar brightness were selected from the frames for comparison: R133, R137, R138, Mk\,11, and R146. From these, differential residuals ($\Delta V$) were calculated for R139 compared to the mean magnitude of the check star for each epoch. The deviation was found to be consistent with that calculated between the check stars themselves, indicating that R139 shows no photometric variability. These results are shown in Figure \ref{phot}, where the observations have been phased to the 153.9\,day orbit. 

If the inclination ({\it i}) of the system was $90^{\circ}$, the maximum duration of eclipses near apastron and periastron has been calculated to be 7.9 and 2.9\,days respectively. From the sampling of our photometric data, it is unlikely that such events would have gone undetected, see Figure \ref{phot}. However, if the inclination is lower ($80^{\circ}\!\!\lesssim\!i\! \lesssim \!\!86^{\circ}$), there is no eclipse near apastron and the periastron eclipse is shorter. Consequently, an intensive photometric observing campaign is required near to periastron to conclusively determine if there is any evidence for an eclipse.

%==========================================================================
%==========================================================================
%

\section{X-rays}\label{xrays}
R139 was detected by \cite{zwart02} as an X-ray source in the 30~Dor field observed with the Advanced CCD Imaging Spectrometer on the \emph{Chandra X-Ray Observatory}. Further analysis of the \emph{Chandra} data was carried out by \citet{townsley} and more recently by \citet{guerrero}. These studies found R139 to have a relatively low X-ray luminosity compared with other W-R stars in the region. \citet{guerrero} also considered data from the R\"ontgen Satellite (\emph{ROSAT}), but it did not detect R139 owing to its lower sensitivity.

The X-ray luminosity and the bolometric luminosity of massive O stars are linked by the relationship $L_X \approx 10^{-6.9}L_{bol}$ \citep{sana06X}. Therefore, with a luminosity of $\log(L/L_{\odot})=6.4$ for the combined system, an X-ray luminosity of $1.2 \times 10^{33} \, \rm{erg\,s^{-1}}$ would be expected. This is slightly lower than Guerrero's result of $2.7 \times 10^{33} \, \rm{erg\,s^{-1}}$ in the 0.5 - 7.0keV range - even considering the possible 25\% error in the detected count rate. This slight excess emission might be associated with X-rays generated through the interaction of the system's stellar winds. There is no evidence, however, for phase-dependent line profile variations, which would have also suggested colliding winds.

%==========================================================================
%==========================================================================
%

\section{Discussion}\label{discussion} 
\emph{Previous observations:} In order to compare our result with the earlier work of \citet{schnurr}, who found an insignificant radial velocity shift, our LR02 observations were degraded to the same resolving power as Schnurr's ({\it R}\,$\sim1000$) and a number of lines were fit with a single Gaussian function. The radial velocity variation was found to be only 10.3\,km\,s$^{-1}$, while the FWHM of the profiles varied by around 40\,km\,s$^{-1}$. This suggests that, even if the system had been observed near periastron, it would have been difficult to confirm its binary nature. 
 
\vspace{0.2cm}
\noindent{\emph{Comparison systems:}\ 
%\ \newline
Some binary systems have been identified where both components are more massive than those of R139: NGC-3603-A1, is a system comprised of a 116M$_{\sun }$ primary and a 89M$_{\sun }$ star \citep{schnurr_binary}, and also WR20a, an 83M$_{\sun }$ and 82M$_{\sun }$ system \citep{wr20_rauw,wr20_bonanos}. However, there are not many systems with a pair of massive evolved O-stars. Closer analogs are the Cyg OB2-B17 system \citep{vanessa}, where the component stars are O7 and O9 supergiants and Cyg OB2-\#5 where one of the stars is an O6-7 supergiant \citep{rauw99}. It would appear that neither of these systems contain stars as massive as those predicted here. Consequently, it can be argued that R139 is the most massive O supergiant binary system yet discovered. }

\begin{figure}[!t]
\centering
\includegraphics[width=7.5cm,height=4.8cm]{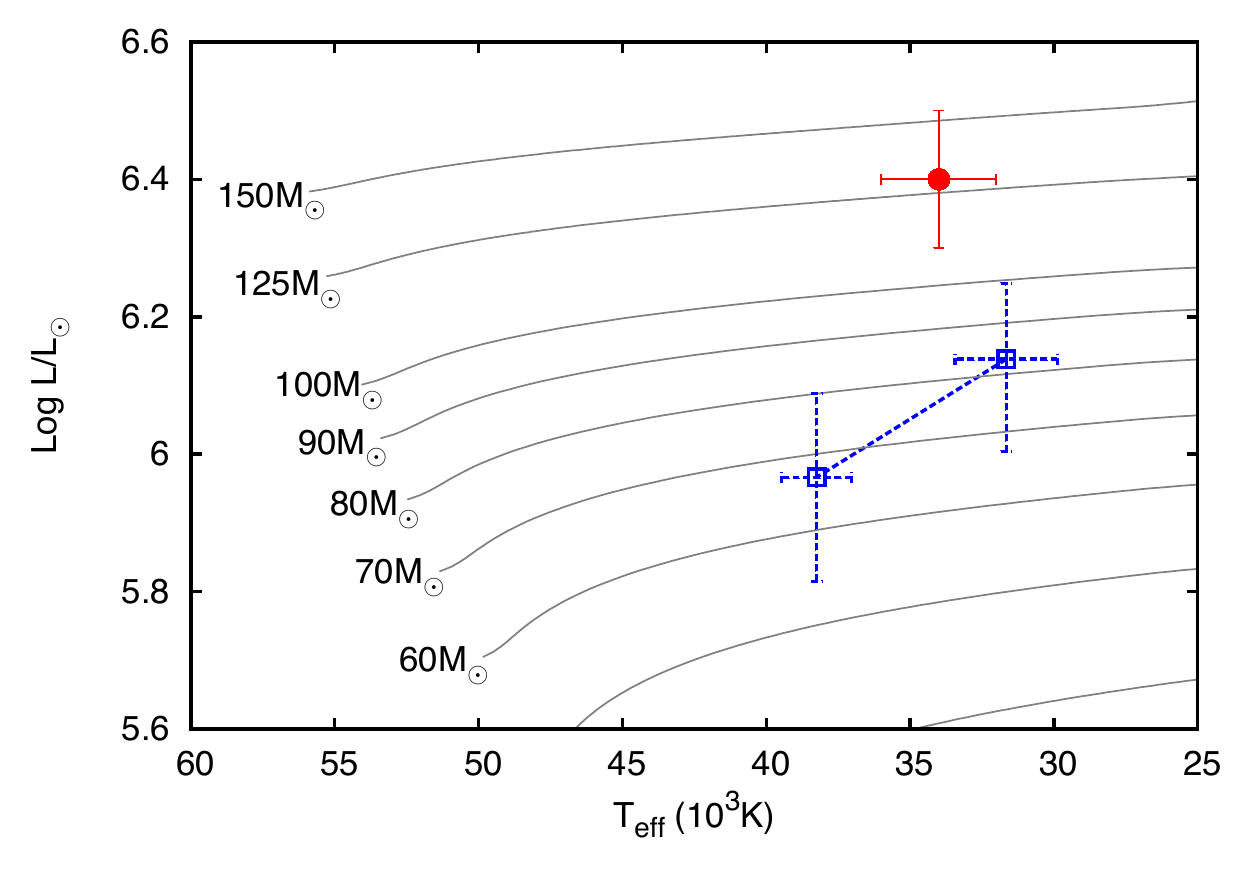}
\caption{Hertzsprung-Russell diagram showing mass estimates derived from luminosity fits to evolutionary tracks (Brott et al. 2011; Friedrich et al. in prep.).~The red circle indicates the best fit for the total luminosity of the R139 system, while the blue squares show the fit for the luminosity of the two components based on the mass ratio and assuming the objects are coeval. From these it is possible to infer the initial masses, see text for further details.  
\label{hrd}}
\vspace{-0.3cm}
\end{figure}

\vspace{0.2cm}
\noindent{\emph{Evolutionary masses:}\
Figure \ref{hrd} shows how the R139 system compares to evolutionary tracks from Friedrich et al. (in prep.), computed analogously to the models of \cite{ines}. The figure shows that the total luminosity and $T_{\rm eff}$ of the system equals that of a single star with an initial mass above 125 $M_{\sun}$.}

Assuming R139 consists of coeval stars with a mass ratio of 1.2,  the estimates for the current masses are $75 \pm 14 M_{\sun}$ for the (cooler) primary and $62 \pm 11 M_{\sun}$ for the (hotter) secondary.  These values have been derived using a $\chi^2$ method to fit the combined luminosity of the stars against that quoted for the system.  The effective temperatures of the stars were fitted against the CMFGEN-derived temperatures of $34 \pm 2$\,kK. Interestingly, these estimated masses are in close agreement with the lower-mass limits from the orbital solution. This implies that the system has a high inclination and supports the need for additional photometric observations.

In these models an initial equatorial velocity of 110 km\,s$^{-1}$ was adopted in agreement with the current observed $v \sin i$. The effect of rotation on the evolutionary tracks is very limited for initial rotation rates up to about 300\,km\,s$^{-1}$\citep{ines}. Nevertheless, these tracks are sensitive to uncertain physical processes such as internal mixing and mass loss. The errors on the mass estimates represent the formal 1-sigma confidence limits of the $\chi^2$ fit and do not include systematic uncertainties in the model physics.

The best fit corresponds to an age of 2 - 2.5\,Myr and implies that both stars have significantly evolved off the zero-age main sequence. As the stars are assumed to be coeval, the substantial mass ratio implies a large difference in temperature between the components (see Fig. \ref{hrd}). This is surprising given the similar spectral types; \citet{martins} predict a temperature difference nearer to 1\,kK for a 0.5 variation in spectral types. This discrepancy may well reflect our still limited understanding of the physics of the most massive stars, illustrating the potential of massive binaries as tools to evaluate our models.

\vspace{0.4cm}
\noindent{The high quality, time-sampled VFTS observations have revealed that R139 is a binary system. The data suggest that it is the most massive evolved O-star binary system yet discovered: a result which further observations around periastron would help to confirm. As demonstrated here, such a massive system has already presented challenges for theoretical models to reproduce its observed properties and it will likely provide a crucial test for evolutionary and atmospheric models in the future.
}

%==========================================================================
%
\acknowledgements{Thanks to the referee, Anthony Moffat, for their constructive comments. SdM acknowledges NASA Hubble Fellowship grant HST-HF-51270.01-A awarded by STScI, operated by AURA for NASA, contract NAS 5-26555. AZB acknowledges support from the European Commission\,FP7 under the Marie Curie International Reintegration Grant PIRG04-GA-2008-239335. RHB acknowledges partial support from DIULS Project PR09101.}

\bibliographystyle{aa} 
\bibliography{r139_bib2}

\end{document}